\documentclass[journal]{IEEEtran}
\usepackage{amsmath,amsfonts}
\usepackage{algorithmic}
\usepackage{algorithm}
\usepackage{array}
\usepackage{bm}
\usepackage{textcomp}
\usepackage{stfloats}
\usepackage{url}
\usepackage{verbatim}
\usepackage{graphicx}
\usepackage{cite}
\usepackage{amssymb}
\usepackage{CJK}
\usepackage{indentfirst}
\usepackage{cases}

\usepackage{setspace}
\usepackage{hyperref}
\usepackage{float}
\usepackage{titlecaps}
\usepackage{acronym}
\usepackage{xcolor}

\acrodef{isac}[ISAC]{integrated sensing and communications}
\acrodef{sinr}[SINR]{signal-to-interference-plus-noise ratio}
\acrodef{snr}[SNR]{signal-to-noise ratio}
\acrodef{crb}[CRB]{Cram{\'e}r-Rao Bound}
\acrodef{dof}[DoF]{degrees of freedom}
\acrodef{miso}[MISO]{ multiple-input single-output}
\acrodef{mimo}[MIMO]{multiple-input and multiple-output}
\acrodef{mu-mimo}[MU-MIMO]{multi-user multiple-input and multiple-output}
\acrodef{mi}[MI]{mutual information}
\acrodef{pwm}[PWM]{planar-wave model}
\acrodef{swm}[SWM]{spherical-wave model}
\acrodef{nusw}[NUSW]{non-uniform spherical-wave}
\acrodef{hspm}[HSPM]{hybrid spherical and planar wave model}
\acrodef{aoa}[AoA]{angle of arrival}
\acrodef{aod}[AoD]{angle of departure}
\acrodef{hbf}[HBF]{hybrid beamforming}
\acrodef{los}[LoS]{line-of sight}
\acrodef{nlos}[NLoS]{non-line-of sight}
\acrodef{awgn}[AWGN]{additive white Gaussian noise}
\acrodef{rcs}[RCS]{radar cross section}
\acrodef{fim}[FIM]{Fisher’s information matrix}
\acrodef{xl-mimo}[XL-MIMO]{extremely large-scale MIMO}
\acrodef{mmwave}[mmWave]{millimeter wave}
\acrodef{xl}[XL]{extremely large-scale}
\acrodef{em}[EM]{electromagnetic}
\acrodef{iid}[i.i.d.]{independent and identically distributed}
\acrodef{rx}[Rx]{receiver}
\acrodef{tx}[Tx]{transmitter}
\acrodef{bs}[BS]{base station}

\ifCLASSOPTIONcompsoc
\usepackage[caption=false,font=normalsize,labelfont=sf,textfont=sf]{subfig}
\else
\usepackage[caption=false,font=footnotesize]{subfig}
\fi

\hypersetup{colorlinks=true}
\def\BibTeX{{\rm B\kern-.05em{\sc i\kern-.025em b}\kern-.08em
		T\kern-.1667em\lower.7ex\hbox{E}\kern-.125emX}}

\newtheorem{theorem}{\textbf{Theorem}}

\newtheorem{lemma}{\textbf{Lemma}}

\newtheorem{corollary}{\textbf{Corollary}}

\newtheorem{proposition}{\textbf{Proposition}}

\makeatletter

\newcommand{\Rmnum}[1]{\expandafter\@slowromancap\romannumeral #1@}
\makeatother

\usepackage{framed} 

\usepackage{cancel}

\begin{document}

	\title{Cram{\'e}r-Rao Bounds for Near-Field Sensing: A Generic Modular Architecture}
	\author{Chunwei~Meng, Dingyou~Ma,  Xu~Chen, Zhiyong~Feng,  and~Yuanwei~Liu
		\thanks{C. Meng, D. Ma, X. Chen and Z. Feng are with the Key Laboratory of Universal Wireless Communications, Ministry of Education, Beijing University of Posts and Telecommunications, Beijing 100876, China (e-mail:
			mengchunwei@bupt.edu.cn;
			dingyouma@bupt.edu.cn;
			 chenxu96330@bupt.edu.cn;
			fengzy@bupt.edu.cn). 
		Yuanwei~Liu is	with the School of Electronic Engineering and Computer Science, Queen Mary University of London, London E1 4NS, U.K. (e-mail:  yuanwei.liu@qmul.ac.uk).	
		}
	}
	\maketitle
	
	\begin{abstract}
    A generic modular array architecture is proposed, featuring uniform/non-uniform subarray layouts that allows for flexible deployment.
	The bistatic near-field sensing system is considered, where the target is located in the near-field of the whole modular array and the far-field of each subarray.
	Then, the closed-form expressions of Cramér-Rao bounds (CRBs) for range and angle estimations are derived based on the  hybrid spherical and planar wave model (HSPM).
	Simulation results validate the accuracy of the derived closed-form CRBs and demonstrate that: i) The HSPM with varying angles of arrival (AoAs) between subarrays can  reduce the CRB for range estimation compared to the traditional HSPM with shared AoA;
    and	ii)  The proposed generic modular architecture with subarrays positioned closer to the edges can significantly reduce the CRBs compared to the traditional modular architecture with uniform subarray layout, when the array aperture is fixed.
		
	\end{abstract}
	

\section{Introduction}

\IEEEPARstart{I}{n} the sixth generation wireless systems, the \ac{xl} antenna arrays and \ac{mmwave}/terahertz (THz) are considered key technologies to  significantly enhance the spectrum efficiency and spatial resolution, which is essential to meet the demands for high-capacity communications and high-resolution sensing of emerging applications such as smart manufacturing and smart home\cite{liu2023tutorial,wang2023nearisac}. 
Nevertheless, the larger array apertures and higher frequency bands will result in communication users or sensing targets being located in the near-field region, rendering the traditional far-field channel model based on planar electromagnetic wavefront invalid\cite{lu2023tutorial}. 
Therefore, it is necessary to consider more general spherical wavefront characteristics for both near-field communications and sensing.

Recently, the modular \ac{xl}-array has been regarded as a promosing architecture utilized for the \ac{mmwave}/THz frequency bands, which is comprised of multiple identical subarrays with uniform array antennas and relatively large intervals between adjacent subarrays\cite{lu2023tutorial,yan2021joint,li2023multi}.
The modular XL-array with a larger array aperture enables an extended near-field range, improved spatial resolution, and flexible deployment\cite{li2022nearfield}.
Therefore, \cite{yang2023performance} investigated the potential of modular \ac{xl}-arrays  in near-field sensing and conducted a detailed analysis of the range and angle \ac{crb}s.
However, the studies mentioned above considered the modular \ac{xl}-array with uniformly arranged subarrays, which can result in unignorable grating lobes\cite{li2023multi}. 
In \cite{Kao2023fast}, a novel approach is proposed based on a non-uniform subarray layout to suppress the sidelobes of the modular \ac{xl}-array and enable ambiguity-free \ac{aoa} estimation.
The aurthors in \cite{Khami2021condi} derived \ac{crb}s based on \ac{swm} with antenna arrays of arbitrary geometry. 
However, in a modular architecture, due to the smaller subarray aperture compared to the entire array aperture, the target is more likely to be located in the near-field of the entire array and the far-field of each subarray\cite{yan2021joint}.  
Therefore, employing the \ac{hspm}  is more suitable due to its advantages in low modeling and computational complexity over the \ac{swm}.

In this letter, we propose a novel generic modular array architecture with   uniform/non-uniform subarray arrangement and investigate its potential in near-field sensing.
We consider a bistatic near-field sensing system operating in the phased-array radar mode.
Additionally, we derive closed-form expressions of the range and angle \ac{crb}s based on the \ac{hspm} model, considering varying subarray \ac{aoa}s.
Simulation results demonstrate that: 1) The \ac{hspm} with distinct \ac{aoa}s yields significantly lower\ range \ac{crb}  compared to the commonly used \ac{hspm} with shared \ac{aoa}, and closely approximates  the \ac{swm} in terms of the range and angle \ac{crb}s. 
2) In a centro-symmetric modular array with a fixed aperture, a non-uniform arrangement of subarrays near the edges achieves lower range and angle \ac{crb}s than a traditional uniform subarray arrangement. 
This validates the effectiveness of the proposed generic modular array architecture in near-field sensing.

\section{System Model}

\begin{figure} [t]
		\setlength{\abovecaptionskip}{-0.2cm}  
		\centering
		\includegraphics[width=0.18\textwidth, height=0.14\textheight]{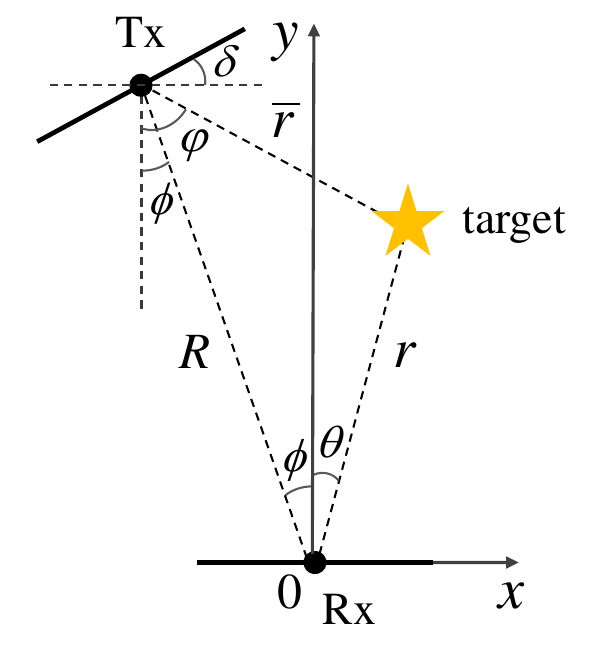}
		\caption{\textcolor{black}{Illustration of the bistatic near-field sensing system.}}
		\label{system_model}
		\setlength{\belowcaptionskip}{-1cm} 
	\end{figure}

We consider a bistatic near-field sensing system as illustrated in Fig.~\ref{system_model}, with application scenarios including unmanned aerial vehicles surveillance and assisted driving in intelligent transportation systems\cite{zhang2024pmn,xiang2022vcn}.
	The  \ac{tx} is equipped with a uniform \ac{xl}-array  comprising $N_t$ antennas.
\textcolor{black}{	The \ac{rx}  is equipped with a generic modular \ac{xl}-array  incorporating $N_r=KM$ antennas, where $K$ is the number of subarrays and $M$ is the number of antenna elements within each subarray.
   The distance between the centers of \ac{tx} and \ac{rx} is $R$.}
   The inter-antenna spacing of \ac{tx}/\ac{rx} is denoted as $d=\frac{\lambda}{2}$, where $\lambda$ denotes the wavelength.
    For notational convinience, we assume that $N_t$, $K$, and $M$ are odd numbers, so that the $n$-th antenna at \ac{tx}, the $k$-th subarray at \ac{rx}, and  $m$-th antenna of each subarray belong to the integer sets $\mathcal{N}={\left\{0, \pm 1, \cdots, \pm \frac{N_t-1}{2}\right\}}$,  $\mathcal{K}={\left\{0, \pm 1, \cdots, \pm \frac{K-1}{2}\right\}}$, and $\mathcal{M}={\left\{0, \pm 1, \cdots, \pm \frac{M-1}{2}\right\}}$, respectively.

\textcolor{black}{	Without loss of generality, we assume that the generic modular \ac{xl}-array of \ac{rx} is placed along the $x$-axis, while the \ac{tx} array is oriented at an angle $\delta$ with respect to the \ac{rx} array.}
	For each subarray, we select its center element as the reference antenna.
	\textcolor{black}{The distance between the k-th subarray and its adjacent subarray closer to the origin is $\Gamma_kd$, where $\Gamma_k\geq1$ is an integer determined by the practical deployment conditions.}
	Therefore,  the coordinate of the $m$-th element of the $k$-th subarray can be represented as $\mathbf{l}_{k, m}=(x_{k,m},0)$, where $x_{k,m} = (\sum\nolimits_{i=0}^{k}{{{\Gamma }_{i}}}+k(M-1)+m)d$ for $k>0$, and $x_{k,m}= (-\sum\nolimits_{i=0}^{k}{{{\Gamma }_{i}}}+k(M-1)+m)d$ for $k\leq0$.
	%
	\begin{figure} [t]
		\setlength{\abovecaptionskip}{-0.2cm} 
		\setlength{\belowcaptionskip}{-1cm} 
		\centering
		\includegraphics[width=0.49 \textwidth, height=0.13\textheight]{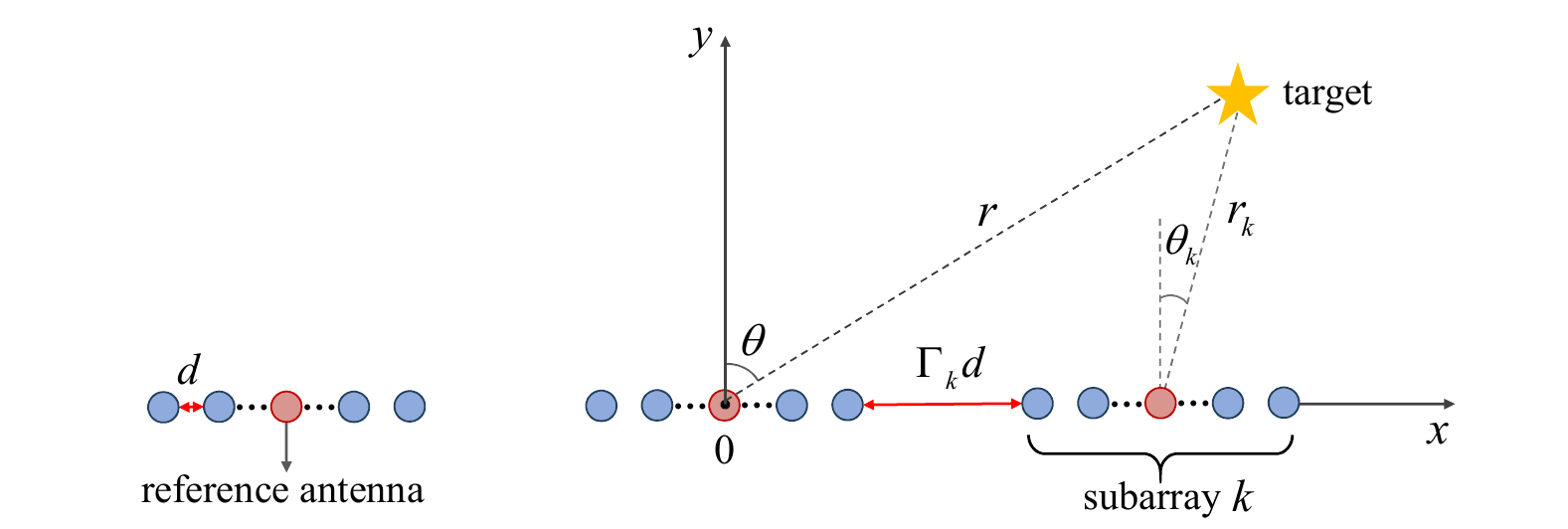}
		\caption{The generic modular array with $K$ subarrays and $M$ antennas in each subarray.}
		\label{array_configuration_fig}  
	\end{figure}

	The radar target is located at $\mathbf{l}_{q}=(r \sin \theta, r \cos \theta)$, where $r$ is its distance from the origin, and $\theta \in\left[-\frac{\pi}{2}, \frac{\pi}{2}\right]$ is its angle with respect to the positive $y$-axis.
   Let $\varphi$, $\phi$, and $\bar{r}$ denote the \ac{aoa}, the relative angle, and distance between the target and  the center of \ac{tx} array, respectively.
    \textcolor{black}{With prior knowledge about the relative positions of Tx and Rx, specifically the known parameters $R$, $\phi$, and $\delta$, distance $\bar{r}$ and angle $\varphi$ at \ac{tx} can be expressed in terms of the parameters $r$ and $\theta$ at \ac{rx}, namely,}
    \begin{equation}
    \setlength{\abovedisplayskip}{2pt}
	\setlength{\belowdisplayskip}{2pt}
    	\begin{aligned}
    		&\bar{r}(r,\theta) =\sqrt{{{R}^{2}}+{{r}^{2}}-2Rr\cos (\theta +\phi )},\\
    		&\varphi(r,\theta) =\arcsin \left\{ \frac{r\sin\theta+R\sin\phi}{\sqrt{{{R}^{2}}+{{r}^{2}}-2Rr\cos (\theta +\phi )}} \right\}.
    	\end{aligned}
    \end{equation}
    \textcolor{black}{In cases where $\bar{r}\leq\frac{2[(N_t-1)d]^2}{\lambda}$, the transmit array response vector based on \ac{swm} is given by  ${\mathbf{g}_{t}}(r,\theta)={{[ e{{}^{-j\frac{2\pi }{\lambda }\bar{r}_n}} ]}_{\forall n\in \mathcal{N}}}$, with $	\bar{r}_n  =\sqrt{{\bar{r}}^2-2 \bar{r} nd \sin (\varphi-\delta) +(nd)^2}$.}

	Note that the large inter-subarray spacing of Rx, i.e., $\Gamma_{k} d$, results in an  expanded array aperture, which is given by $S= (\sum\nolimits_{k\in\mathcal{K}}{{{\Gamma }_{i}}}+K(M-1))d$.
	When $ \frac{2[(M-1)d]^2}{\lambda} \leq r \leq \frac{2S^2}{\lambda} $, the target is located in the near-field of the entire array and the far-field of individual subarrays.
   Then, the \ac{hspm} becomes applicable by combining spherical wave propagation between subarrays and planar wave propagation for each subarray\cite{yan2021joint,Chen2021Hybrid}.
	Specifically, the distance  between the target and  the reference antenna of the $k$-th subarray in \ac{rx} can be expressed as
	\begin{equation}
		r_k =\left\|\mathbf{l}_q-\mathbf{l}_{k,0}\right\|  =\sqrt{r^2-2 r x_k \sin \theta+(x_k)^2},
	\end{equation}
	where $x_k \triangleq x_{k,0}$ denotes the abscissa cooridnate of  the $k$-th subarray's reference antenna.
	By utilizing the \ac{hspm} with distinct \ac{aoa}s\cite{li2023multi}, the \ac{rx} array response vectors can be obtained as 
	\begin{equation}
		{{\mathbf{g}}_r}=\left( \text{diag}\left( {{\boldsymbol{\nu }}} \right)\otimes {{\mathbf{I}}_{M}} \right){{{\mathbf{\tilde{a}}}}},
	\end{equation}
	where  ${{\boldsymbol{\nu}}}={{[ e{{}^{-j\frac{2\pi }{\lambda }r_k}} ]}_{\forall k\in \mathcal{K}}}$ represents the inter-subarray response vector under the uniform spherical-wave assumption, ${{\mathbf{\tilde{a}}}}={{[ {{( \mathbf{a}_{1} )}^{T}},\ldots ,{{( \mathbf{a}_{K} )}^{T}}]}^{T}}\in \mathbb{C}^{KM \times 1}$ with ${{\mathbf{a}}_{k}}={{[{{e}^{j\frac{2\pi }{\lambda }\frac{(M-1)}{2}d\sin {{\theta }_{k}}}},\ldots ,{{e}^{-j\frac{2\pi }{\lambda }\frac{(M-1)}{2}d\sin {{\theta }_{k}}}}]}^{T}}, \forall k\in\mathcal{K}$  denotes the intra-subarray response vectors under the uniform planar-wave assumption.
	The angle of target observed from the $k$-th subarray's reference antenna is denoted as  ${\theta}_{k}$, which can be  obtained as
	\begin{equation}\label{sin_thetak}
 \setlength{\abovedisplayskip}{2pt}
					\setlength{\belowdisplayskip}{2pt}
		\sin \theta _{k}=\frac{r\sin \theta -{{x}_{k}}}{\sqrt{r^2-2 r x_k \sin \theta+(x_k)^2}}.
	\end{equation}

In order to track the target at a specific position, the \ac{tx} operates in a phased-array radar mode, generating the transmit beam towards the specific distance $r^{\prime}$ and angle ${\theta}^{\prime}$.
\textcolor{black}{Within the coherent processing interval, the target is assumed to be stationary.
Consequently, the narrow-band transmitted signal is given by}
\begin{equation}\label{transmit_signal}
	\mathbf{x}\left( t \right)=\mathbf{w}_ts\left( t \right),
\end{equation}
\textcolor{black}{where $\mathbf{w}_t=\mathbf{g}_t^*(\bar{r},\phi-\delta)\in\mathbb{C}^{N_t\times1}$ is the transmit beamfocusing vector towards the target at the position $(\bar{r},\phi-\delta)$ in the Tx-centric coordinate system, which corresponds to the target position $(r',\theta')$ in the Rx-centric coordinate system, and $s(t)$ represents the transmitted waveform.}
\textcolor{black}{The \ac{rx} employs fast-time sampling within each pulse to obtain the received signal. }
 Therefore, the received signal for bistatic near-field sensing is 
	\begin{equation} \label{receive_signal}
	\mathbf{r}\left( t \right)=\beta {{\mathbf{g}}_{r}}\mathbf{g}_{t}^{T}\mathbf{x}\left( t-\tau \right)+\mathbf{n}\left( t \right),
	\end{equation}
	where $\beta$ is  the complex reflection coefficient proportional to the \ac{rcs} of the target, \textcolor{black}{ $\tau=\frac{\bar{r}+r}{c}$ is the round-trip time delay from the transmitter to the target and back to the center antenna of \ac{rx}, $c$ is the speed of light, }
	and $\mathbf{n}(t)$ denotes the \ac{awgn}, with each element being \ac{iid} circularly symmetric complex Gaussian noise with power spectral density $\sigma_s^2$.
Subsequently, by substituting (\ref{transmit_signal}) into (\ref{receive_signal}) and performing matched filtering, we can obtain
		\begin{equation}
			\begin{aligned}
				 \mathbf{y}&=\frac{1}{T}\int_{T}{\mathbf{r}}(t){{s}^{*}}(t-\zeta )dt \\ 
				& =\beta {{\mathbf{g}}_{r}}\mathbf{g}_{t}^{T}(r,\theta )\mathbf{g}_{t}^{*}\left( {r}',{\theta }' \right)R\left( \zeta -\tau  \right)+\mathbf{\tilde{n}},  
			\end{aligned}	
		\end{equation}
	where $R\left( \zeta -\tau  \right)=\frac{1}{T}\int_{T}{s(t-\tau ){{s}^{*}}(t-\zeta )}dt$ is the autocorrelation function, and
	$\mathbf{\tilde{n}}$ is the noise vector obtained after matched filtering, with zero mean and variance $\sigma_s^2$.
	When the parameters of both the matched filter and the transmit beamforming align with the true values of the target parameters,  i.e., $r'=r$, ${\theta}'=\theta$, $ \zeta =\tau$, the transmit beam can be focused towards the target, thereby achieving maximized \ac{sinr} at \ac{rx}\cite{wang2022snr}.
	Therefore, the output of the received signal after matched filtering is given by 
	\begin{equation}
		\mathbf{y}=\alpha \mathbf{g}_r+\mathbf{\tilde{n}},
	\end{equation}
	where $\alpha \triangleq \beta N_t$ denotes the  constant coefficient gain.

	\section{Performance Analysis of Cram{\'e}r-Rao Bound}\label{perform_crb}
	
	Let $\boldsymbol{\eta }={{\left[ r,\theta ,{{\beta }_{r}},{{\beta }_{i}} \right]}^{T}}$ represent the vector that incorporates all unknown target-related parameters, with ${{\beta }_{r}}$ and ${{\beta }_{i}}$ denoting the real and imaginary parts of $\beta$, respectively.
	Since the spherical wavefront curvature contains  two-dimensional  angle and distance information, a direct positioning estimation method can be employed to extract the target's position information from the received signal\cite{lu2022nfchannel,hua2023near}.
	In this letter, we use \ac{crb} to evaluate the performance of the target position estimation.

	Following, we derive the  closed-form  \ac{crb}s of the generic modular XL-array under four different wavefront assumptions, namely: 1) \ac{hspm} with distinct \ac{aoa}s, 2) \ac{hspm} with shared \ac{aoa}, 3) \ac{pwm} , and 4) \ac{swm}.
\textcolor{black}{
	\begin{theorem} \label{crb_r_theta}
		\emph{The  \ac{crb}s  based on \ac{hspm} with distinct \ac{aoa}s for estimating range and angle can be expressed as follows, respectively,
	\begin{subequations} \label{crb_closedform}
		\setlength{\abovedisplayskip}{3pt}
		\setlength{\belowdisplayskip}{2pt}
		\begin{align}
			\label{crb_r_hspm_dis}
			&\text{CRB}_r^{\text{HSPM-Dist}}\notag\\[-4mm]
		&=\frac{\frac{6\sigma_{s}^2K}{{{\left| \alpha  \right|}^{2}}M}{{\left( \frac{\lambda }{2\pi } \right)}^{2}}}{K\!\left(\! {{M}^{2}}\!-\!1\! \right){{d}^{2}}z\!+\!12Kq\!-\!12{{p}^{2}}\!-\!\frac{{{\left[ K\left( {{M}^{2}}-1 \right){{d}^{2}}\hat{z}+12K\hat{q}-12p\tilde{p} \right]}^{2}}}{K\left( {{M}^{2}}-1 \right){{d}^{2}}\tilde{z}+12K\tilde{q}-12{{{\tilde{p}}}^{2}}}},\\[-3mm]
			\label{crb_theta_hspm_dis}
			&\text{CRB}_{\theta}^{\text{HSPM-Dist}}\notag\\[-4mm]
			&=\frac{\frac{6\sigma_{s}^2K}{{{\left| \alpha  \right|}^{2}}M}{{\left( \frac{\lambda }{2\pi } \right)}^{2}}}{K\!\left(\! {{M}^{2}}\!-\!1\! \right){{d}^{2}}\tilde{z}\!+\!12K\tilde{q}\!-\!12{{{\tilde{p}}}^{2}}\!-\!\frac{{{\left[ K\left( {{M}^{2}}-1 \right){{d}^{2}}\hat{z}+12K\hat{q}-12p\tilde{p} \right]}^{2}}}{K\left( {{M}^{2}}-1 \right){{d}^{2}}z+12Kq-12{{p}^{2}}}},
		\end{align}
	\end{subequations}
			where $p\triangleq\sum\nolimits_{k\in \mathcal{K}}{\frac{\partial {{r}_{k}}}{\partial r}}$, $\tilde{p}\triangleq\sum\nolimits_{k\in \mathcal{K}}{\frac{\partial {{r}_{k}}}{\partial \theta }}$, $q\triangleq{{\sum\nolimits_{k\in \mathcal{K}}{\left( \frac{\partial {{r}_{k}}}{\partial r} \right)}}^{2}}$, $\tilde{q}\triangleq\sum\nolimits_{k\in \mathcal{K}}{{{\left( \frac{\partial {{r}_{k}}}{\partial \theta } \right)}^{2}}}$, $\hat{q}\triangleq\sum\nolimits_{k\in \mathcal{K}}{\left( \frac{\partial {{r}_{k}}}{\partial r} \right)\left( \frac{\partial {{r}_{k}}}{\partial \theta } \right)}$,
			$z\triangleq\sum\nolimits_{k\in \mathcal{K}}{{{\left( \frac{\partial \sin {{\theta }_{k}}}{\partial r} \right)}^{2}}}$,
			$ \tilde{z}\triangleq\sum\nolimits_{k\in \mathcal{K}}{{{\left( \frac{\partial \sin {{\theta }_{k}}}{\partial \theta } \right)}^{2}}}$, and$\hat{z}\triangleq\sum\nolimits_{k\in \mathcal{K}}{\left( \frac{\partial \sin {{\theta }_{k}}}{\partial r} \right)\left( \frac{\partial \sin {{\theta }_{k}}}{\partial \theta } \right)}$. 
   The corresponding derivatives in the above expressions are given by
   \begin{equation} \label{theorem1_inter}
       \begin{aligned}
  & \frac{\partial {{r}_{k}}}{\partial r}\!\!=\!\!\frac{r-{{x}_{k}}\sin \theta }{\sqrt{{{r}^{2}}\!-\!2r{{x}_{k}}\sin \theta\! +\!{x}_{k}^{2}}}, \frac{\partial \sin\!{\theta}_k}{\partial r}\!\!=\!\!\frac{r{{x}_{k}}{{\cos }^{2}}\theta }{{\left( {{r}^{2}}\!-\!2r{{x}_{k}}\sin \theta \!+\!{x}_{k}^{2} \right)}^{\frac{3}{2}}},  \\ 
 & \frac{\partial {{r}_{k}}}{\partial \theta }\!\!=\!\!\frac{-r{{x}_{k}}\cos \theta }{\sqrt{{{r}^{2}}\!-\!2r{{x}_{k}}\sin \theta \!+\!{x}_{k}^{2}}},  \frac{\partial \sin\!{{\theta }_{k}}}{\partial \theta }\!\!=\!\!\frac{{{r}^{2}}\cos \theta \left( r\!-\!\sin \theta {{x}_{k}} \right)}{{\left( {{r}^{2}}\!-\!2r{{x}_{k}}\sin \theta\! +\!{x}_{k}^{2} \right)}^{\frac{3}{2}}}. 
\end{aligned}
   \end{equation}
		}
	\end{theorem}
}
	\begin{IEEEproof}
		Please refer to Appendix \ref{crb_proof}.
	\end{IEEEproof}
  
From (\ref{crb_closedform}), we observe that the impact of different \ac{aoa}s for each subarray on the \ac{crb}s manifests in $z$, $\tilde{z}$, and $\hat{z}$.
\textcolor{black}{It is also noted that both range and angle \ac{crb} expressions contain distance-angle coupling terms, which are characterized by the intermediate variables $\hat{z}$, $\hat{q}$, and $p\tilde{p}$. 
These coupling terms originate from the off-diagonal elements of the Fisher information matrix, indicating the correlation between the estimation of distance and angle parameters.
}

  In contrast, for \ac{hspm} with shared \ac{aoa},  the \ac{aoa}s of different subarrays are considered approximately equal, i.e., $\sin\theta_k=\sin\theta,\forall k$, thus we have $z=0$, $\tilde{z}=K\cos^2\theta$, and $\hat{z}=0$.
  Consequently, we can readily derive the closed-form \ac{crb}s under the \ac{hspm} with shared \ac{aoa} from Theorem \ref{crb_r_theta}, as follows.
	\begin{proposition} \label{hspm_shared_crb}
	    \emph{The  \ac{crb}s based on \ac{hspm} with shared \ac{aoa}  for estimating range and angle can be  derived as a degeneration of   (\ref{crb_closedform}), which are expressed as follows, respectively,
				\begin{subequations} \label{crb_hspm_share}
				\setlength{\abovedisplayskip}{1pt}
				\setlength{\belowdisplayskip}{2pt}
				\begin{align}
					\label{crb_r_hspm_share}
					&{\text{CRB}}_r^{\text{HSPM-Shared}}\notag\\[-4mm]
					&=\frac{\frac{6\sigma_{s}^2K}{{{\left| \alpha  \right|}^{2}}M}{{\left( \frac{\lambda }{2\pi } \right)}^{2}}}{\!12Kq\!-\!12{{p}^{2}}\!-\!\frac{{{\left[12K\hat{q}-12p\tilde{p} \right]}^{2}}}{K^2\left( {{M}^{2}}-1 \right){{d}^{2}}{(\cos\theta)}^2+12K\tilde{q}-12{{{\tilde{p}}}^{2}}}},\\
					\label{crb_theta_hspm_share}
					&{\text{CRB}}_{\theta }^{\text{HSPM-Shared}}\notag \\[-4mm]
					&=\!\!\frac{\frac{6\sigma_{s}^2K}{{{\left| \alpha  \right|}^{2}}M}{{\left( \frac{\lambda }{2\pi } \right)}^{2}}}{K^2\!\left(\! {{M}^{2}}\!\!-\!\!1\! \right){{d}^{2}}{(\cos\theta)}^2\!\!+\!\!12K\tilde{q}\!\!-\!\!12{{{\tilde{p}}}^{2}}\!\!-\!\!\frac{{{\left[12K\hat{q}-12p\tilde{p} \right]}^{2}}}{12Kq-12{{p}^{2}}}}.
				\end{align}
\end{subequations}}
	\end{proposition}

\textcolor{black}{
From \eqref{crb_hspm_share}, we can observe that the coupling terms only consist of $\hat{q}$ and $p\tilde{p}$,
indicating a weaker distance-angle coupling compared to the \ac{hspm} with distinct \ac{aoa}s.	
}

Subsequently,  we can similarly obtain the \ac{crb}s of range and angle under the \ac{pwm} as presented in the following proposition.
   \begin{proposition}
   	\emph{The \ac{crb}s based on \ac{pwm} for estimating range and angle can be expressed as follows, respectively,
   		\begin{equation} \label{crbs_pwm}
   				\setlength{\abovedisplayskip}{3pt}
   			\setlength{\belowdisplayskip}{2pt}
   		\begin{aligned}
   				\text{CRB}_{r}^{\text{PWM}} & \rightarrow +\infty,\\[-3mm]
   			\text{CRB}_{\theta}^{\textit{PWM}}& \!=\!\frac{\frac{6\sigma _{s}^{2}K}{{{\left| \alpha  \right|}^{2}}{{\cos }^{2}}\theta }{{\left( \frac{\lambda }{2\pi } \right)}^{2}}}{12KM\!\sum\limits_{k\in \mathcal{K}}\!{x_{k}^{2}}\!+\!{{K}^{2}}M\left( {{M}^{2}}\!-\!1 \right){{d}^{2}}\!-\!12M{{( \sum\limits_{k\in \mathcal{K}}{{{x}_{k}}})}^{2}}}.
   		\end{aligned}
   	\end{equation}
   }
   \end{proposition}

  It is observed that the \ac{crb} of range in (\ref{crbs_pwm}) tends towards infinity, as the antenna array is incapable of spatially discerning ranges under \ac{pwm} assumption.
 \textcolor{black}{Furthermore, the absence of distance-angle coupling terms in the angle \ac{crb} implies that the angle estimation performance is independent of $r$.}

Then, building upon \cite{wang2023cramerrao}, we derive the \ac{crb}s based on the \ac{swm} which precisely captures the signal phase variations across all array elements.
\textcolor{black}{
	\begin{proposition}\label{swm_crb_pro}
		\emph{The \ac{crb}s based on \ac{swm} for estimating range and angle can be expressed as follows, respectively,
			\begin{subequations} \label{crbs_swm}
					\setlength{\abovedisplayskip}{3pt}
				\setlength{\belowdisplayskip}{2pt}
				\begin{align}
					& \text{CRB}_{r}^{\text{SWM}}=\frac{\frac{\sigma _{s}^{2}KM}{2{{\left| \alpha  \right|}^{2}}}{{\left( \frac{\lambda }{2\pi } \right)}^{2}}}{KM{{\varpi }_{rr}}-{{\left( {{\varpi }_{r}} \right)}^{2}}-\frac{{{\left( KM{{\varpi }_{r\theta }}-{{\varpi }_{r}}{{\varpi }_{\theta }} \right)}^{2}}}{KM{{\varpi }_{\theta \theta }}-{{\left( {{\varpi }_{\theta }} \right)}^{2}}}}, \\
					& \text{CRB}_{\theta }^{\text{SWM}}=\frac{\frac{\sigma _{s}^{2}KM}{2{{\left| \alpha  \right|}^{2}}}{{\left( \frac{\lambda }{2\pi } \right)}^{2}}}{KM{{\varpi }_{\theta \theta }}-{{\left( {{\varpi }_{\theta }} \right)}^{2}}-\frac{{{\left( KM{{\varpi }_{r\theta }}-{{\varpi }_{r}}{{\varpi }_{\theta }} \right)}^{2}}}{KM{{\varpi }_{rr}}-{{\left( {{\varpi }_{r}} \right)}^{2}}}} , 
				\end{align}
			\end{subequations}
			where ${{\varpi }_{r}}\!\triangleq\! \sum\limits_{k,m}{\frac{\partial {{r}_{k,m}}}{\partial r}}$, ${{\varpi }_{\theta }}\triangleq \sum\limits_{k,m}{}\frac{\partial {{r}_{k,m}}}{\partial \theta }$, ${{\varpi }_{rr}}\triangleq \sum\limits_{k,m}{}{{\left( \frac{\partial {{r}_{k,m}}}{\partial r} \right)}^{2}}$, ${{\varpi }_{r\theta }}\triangleq \sum\limits_{k,m}{\left( \frac{\partial {{r}_{k,m}}}{\partial r} \right)\left( \frac{\partial {{r}_{k,m}}}{\partial \theta } \right)}$, ${{\varpi }_{\theta \theta }}\triangleq \sum\limits_{k,m}{{{\left( \frac{\partial {{r}_{k,m}}}{\partial \theta } \right)}^{2}}}$.		
   }
	\end{proposition}
 }

 \textcolor{black}{
Under the \ac{swm} assumption, the coupling terms present in both range and angle CRBs are characterized by ${\varpi}_{r\theta}$ and ${\varpi}r{\varpi}{\theta}$. 
Compared to the \ac{hspm} for distinct \ac{aoa}s, the  \ac{swm} exhibits stronger distance-angle coupling since it considers the unique distance of each antenna element, which inherently contains information about the distance and angle parameters.}

\textcolor{black}{
In traditional estimation theory, coupling between parameters may degrade accuracy due to error propagation\cite{trees2007bayesian}. 
However, in near-field sensing, coupling terms determined by the target's location and wavefront assumption can provide additional information for parameter estimation, thereby facilitating the joint estimation of range and angle\cite{Huang1991near}.
}

\textcolor{black}{
To gain more intuitive insights, we consider a centro-symmetric array and the special case of $\theta=0$, i.e., the target is located on the y-axis. 
This eliminates the complex distance-angle coupling terms, yielding simplified CRB expressions, as elucidated in the following corollaries.
\begin{corollary} \label{crb_specialcase}
    \emph{When $\theta=0$, the \ac{crb}s in Theorem \ref{crb_r_theta} reduce to 
    \begin{subequations}
    	\setlength{\abovedisplayskip}{3pt}
				\setlength{\belowdisplayskip}{2pt}
        \begin{align}
        \label{crb_r_hspm_dis_special}
            & \text{CRB}_{r}^{\text{HSPM-Dist}}\!\!=\!\!\frac{\frac{6\sigma _{s}^{2}K}{{{\left| \alpha  \right|}^{2}}M}{{\left( \frac{\lambda }{2\pi } \right)}^{2}}}{K\left( {{M}^{2}}-1 \right){{d}^{2}}{z}'+12K{q}'-12{{\left( {{p}'} \right)}^{2}}}, \\ 
        \label{crb_theta_hspm_dis_special}
            & \text{CRB}_{\theta }^{\text{HSPM-Dist}}\!\!=\!\!\frac{\frac{6\sigma _{s}^{2}}{{{\left| \alpha  \right|}^{2}}}{{\left( \frac{\lambda }{2\pi } \right)}^{2}}}{M\left( {{M}^{2}}-1 \right){{d}^{2}}{\tilde{z}}'+12M{\tilde{q}}'},
        \end{align}
    \end{subequations}
    where ${p}'\triangleq \sum\nolimits_{k\in\mathcal{K}}\frac{r}{r_k}$, ${q}'\triangleq \sum\nolimits_{k\in\mathcal{K}}\frac{{{r}^{2}}}{{{r}_k^{2}}}$, ${z}'\triangleq \sum\nolimits_{k\in\mathcal{K}}\frac{{{r}^{2}{x_k^2}}}{{{r}_k^{6}}}$, $\tilde{q}'\triangleq \sum\nolimits_{k\in\mathcal{K}}\frac{{{r}^{2}{x_k^2}}}{{{r}_k^{2}}}$, $\tilde{z}'\triangleq \sum\nolimits_{k\in\mathcal{K}}\frac{r^6}{{{r}_k^{6}}}$, with ${{r}_{k}}=\sqrt{{{r}^{2}}+{{x}_{k}^{2}}}, \forall k\in\mathcal{K}$.
    }
\end{corollary}
}
\begin{IEEEproof}
		Please refer to Appendix \ref{corollary_1_proof}.
\end{IEEEproof}

\textcolor{black}{
It can be easily observed from Corollary \ref{crb_specialcase} that as $r$ increases, both $\tilde{z}'$ and $\tilde{q}'$ increase, leading to a decrease in the angle \ac{crb}. 
Moreover, for a fixed number of subarrays $K$, both the range and angle CRBs decrease as the number of antennas within each subarray $M$ increases.
}

\textcolor{black}{
Then, we consider the case where the array aperture is much smaller than the target range, i.e., $\frac{S}{r}\ll 1$, leading to $\frac{x_k}{r}\ll 1$. 
Therefore, $r_k$ can be approximated using the second-order Taylor expansion, i.e., $r_k=\sqrt{{{r}^{2}}+{{x}_{k}^{2}}}\approx r+\frac{{{x}_{k}^{2}}}{2r},\forall k$.
Substituting the approximation of $r_k$ into Corollary \ref{crb_specialcase} yields the following result:
\begin{corollary} \label{crb_asymp}
    \emph{When $\theta=0$ and $\frac{S}{r}\ll 1$, the \ac{crb}s for estimating range and angle in \eqref{crb_r_hspm_dis_special} and \eqref{crb_theta_hspm_dis_special} can be reduced to
    \begin{subequations}
      	\setlength{\abovedisplayskip}{2pt}
				\setlength{\belowdisplayskip}{3pt}
        \begin{align}
            \label{asymp_crb_r}
             & \text{CRB}_{r}^{\text{HSPM-Dist}}\approx\frac{\frac{6\sigma _{s}^{2}K}{{{\left| \alpha  \right|}^{2}}M}{{\left( \frac{\lambda }{2\pi } \right)}^{2}}{{r}^{4}}}{K\left( {{M}^{2}}-1 \right){{d}^{2}}\mathcal{S}_x-3{{\left( \mathcal{S}_x \right)}^{2}}}, \\ 
             \label{asymp_crb_theta}
            & \text{CRB}_{\theta }^{\text{HSPM-Dist}}\approx\frac{\frac{6\sigma _{s}^{2}}{{{\left| \alpha  \right|}^{2}}M}{{\left( \frac{\lambda }{2\pi } \right)}^{2}}}{{{K}}\left( {{M}^{2}}-1 \right){{d}^{2}}+12\mathcal{S}_x-\frac{1}{{{r}^{2}}}\Omega\!\left( {{x}_{k}} \right)},
        \end{align}
    \end{subequations}
    where $\mathcal{S}_x\triangleq\sum\nolimits_{k\in\mathcal{K}}{x_{k}^{2}}$, and $\Omega \left( {{x}_{k}} \right)\triangleq \left( {{M}^{2}}-1 \right){{d}^{2}}\mathcal{S}_x+12\sum\nolimits_{k\in\mathcal{K}}{x_{k}^{4}}$.
    }
\end{corollary}
}

\textcolor{black}{
 It is noted that for the range \ac{crb} in \eqref{asymp_crb_r},  it increases with $r^4$, indicating a rapid deterioration of range estimation accuracy as the target distance increases. 
 Moreover, the denominator term is a quadratic function of $\mathcal{S}_x$, suggesting that as $\mathcal{S}_x$ increases, the range \ac{crb} first decreases and then increases , attaining its minimum value at point $\mathcal{S}_x={K\left( {{M}^{2}}-1 \right){{d}^{2}}}/{6}$.
For the angle \ac{crb} in \eqref{asymp_crb_theta}, when $r$ is relatively large, the first two terms in the denominator dominate.
Consequently, for a fixed array aperture $S$, positioning the subarrays closer to the edges leads to a larger $\mathcal{S}_x$, and thus the lower angle \ac{crb}.
}

	\section{Numerical Results}

	In this section, we validate the theoretical results of range and angle \ac{crb}s based on \ac{hspm} with distinct \ac{aoa}s as derived in Section~\ref{perform_crb}, 
	and investigate the impact of various factors on the \ac{crb}s.
	\textcolor{black}{Unless otherwise specified, the general parameter settings of the system are described as follows:
	the total number of receive antennas is $N=375$, the carrier frequency is $f=60$ GHz, $d={c}/(2f)=0.0025$m\cite{li2023multi}.}
	The set $\mathcal{I}=\{\Gamma_{k}\}_{k\in\mathcal{K}}$ controls the inter-subarray spacing.

\begin{figure} [t]
	\centering
	\includegraphics[width=0.42 \textwidth]{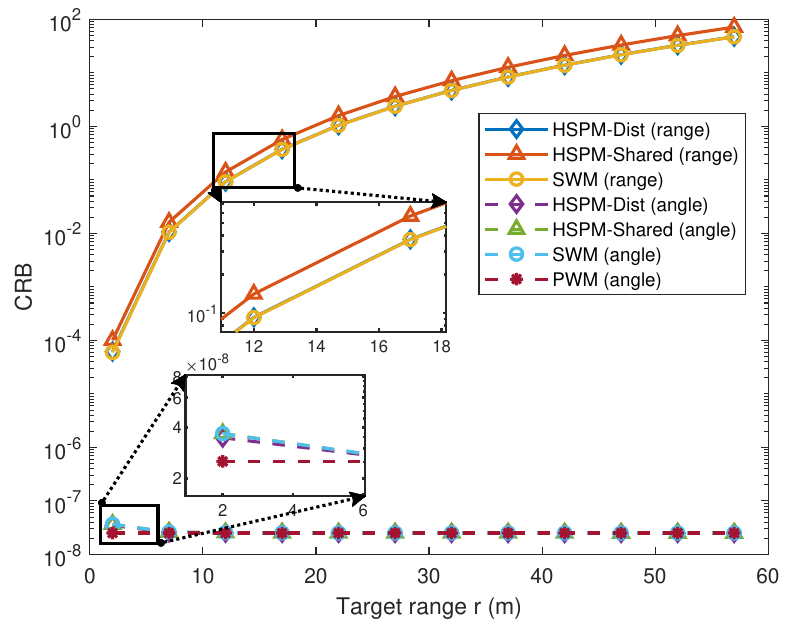}
	\caption{The CRBs of range and angle under different wavefront assumptions
		versus range $r$.}
	\label{CRB_versus_range}  
\end{figure}
    Fig.~ \ref{CRB_versus_range} illustrates the \ac{crb}s for range and angle estimations versus range $r$ for various wavefront assumptions.
    In this case, we set $r\in[1,56]$m, $\theta = \pi/3$, $K=3$ and  $\Gamma_{k}=90$ for all $k\neq0$.
     \textcolor{black}{The \ac{sinr} $\gamma=\frac{|\alpha|^2}{\sigma_s^2}$ is set to $0$ dB\cite{yang2023performance}.}
    %
    The legends ``\ac{hspm}-Dist", ``\ac{hspm}-Shared", ``\ac{swm}" and ``\ac{pwm}" denote the \ac{crb}s for estimating range (solid lines) and angle (dashed lines) calculated as (\ref{crb_closedform}), (\ref{crb_hspm_share}), (\ref{crbs_pwm}) and  (\ref{crbs_swm}), respectively. 
     As $r$ increases, the range \ac{crb} increases for all wavefront assumptions, whereas the angle \ac{crb} exhibits the opposite trend. 
	\textcolor{black}{Specifically, the range CRBs of HSPM-Dist and SWM nearly coincide, while HSPM-Shared exhibits a noticeable discrepancy, indicating that HSPM-Dist closely approaches the accuracy of SWM compared to HSPM-Shared. 
    The improved performance of HSPM-Dist can be attributed to the sine function of each subarray's unique AoA containing additional range information, as demonstrated by (\ref{sin_thetak}).}
    Additionally, noticeable differences are shown among the angle \ac{crb}s under \ac{swm}, \ac{hspm}-Dist, \ac{hspm}-Shared, and \ac{pwm}, indicating notable errors  in the near-field region for \ac{pwm}.
    By contrast,  the angle \ac{crb}s under \ac{hspm}-Dist, \ac{hspm}-Shared, and \ac{pwm} overlap closely within the interval of changing   $r$. 
 \textcolor{black}{
   These observations indicate that the \ac{hspm} with distinct \ac{aoa}s exhibits superior  \ac{crb} performance compared to \ac{hspm} with shared \ac{aoa} and PWM while maintaining lower complexity than SWM. 
   Moreover, in the generic modular XL-array with non-uniformly distributed subarrays, HSPM-Shared's assumption of identical AoAs for all subarrays becomes unreasonable. 
   Consequently, HSPM-Dist strikes a balance between accuracy and complexity, making it the most suitable model for the generic modular XL-array architecture.} 

  Next, we  investigate the impact of different subarray layouts on the range and angle \ac{crb}s  under various wavefront assumptions.
  According to \cite{Gazzah2014crb}, employing a centro-symmetric linear arrays can improve the near-field angle and range estimation capabilities.
  Therefore, we consider   the centro-symmetric modular \ac{xl}-array structure with subarrays arranged in a non-uniform layout symmetric about the origin.
  We fix the positions of subarrays at both ends of the \ac{xl}-array to ensure a constant array aperture. 
    Two subarray configurations, referred to as C1 and C2, are evaluated, with the same total number of antennas but different array apertures. 
   For both configurations, we set $K=5$ and $M=75$.
   By manipulating the position parameters of the two subarrays near the origin, denoted by $\Gamma$, we can control the spacing between subarrays.
    Specifically,  corresponding sets of inter-subarray spacings are defined as  $\mathcal{I}_1=(100-\Gamma, \Gamma, 0,\Gamma,100-\Gamma)$ and $\mathcal{I}_2=(150-\Gamma, \Gamma, 0,\Gamma,150-\Gamma)$ for  C1 and C2, respectively.

\begin{figure} [t]
	\centering
	\includegraphics[width=0.42 \textwidth]{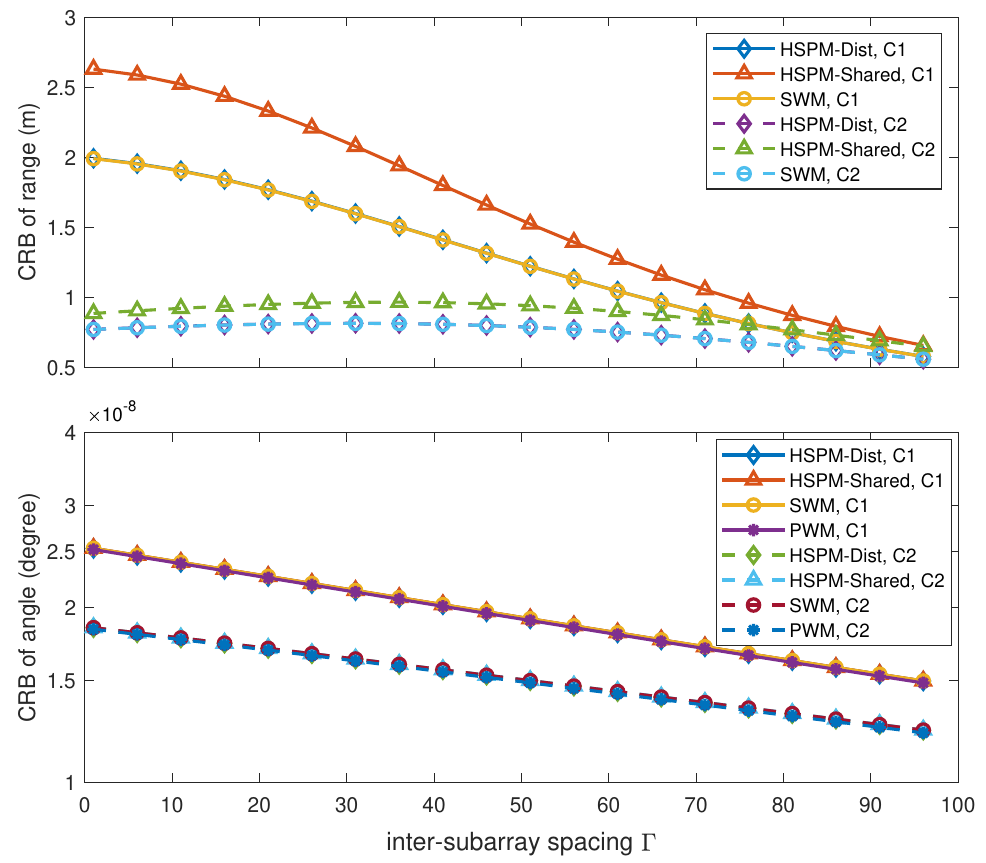}
	\caption{The CRBs of range and angle under different wavefront assumptions
		with different array appertures and subarray layouts.}
	\label{inter_subarray_rtheta}  
\end{figure}

  Fig. \ref{inter_subarray_rtheta} illustrates the changes in  range and angle \ac{crb}s  as $\Gamma$ varies form $1$ to $95$.
  The target distance is $r=30$ m and its angle is $\theta=\pi/3$.
  It can be observed that the angle \ac{crb}s under both  C1 and C2 configurations monotonically decrease as  $\Gamma$ increases.
   However,  the range \ac{crb}s monotonically decrease in C1 configuration , while those under  C2 configuration first increase then decrease in as 
  $\Gamma$ varies.
  When  $\Gamma$ is set to $50$ and $75$, the subarrays in both C1 and C2 are uniformly arranged, respectively.
  Notably, for both C1 and C2, the range and angle \ac{crb}s of the non-uniform subarray arrangements are lower than those of the  uniformly arranged subarrays when $\Gamma$ is greater than 50 and 75, respectively.
  This implies that the modular architecture with uniform subarray layouts is not always the most favorable configuration for near-field sensing. 
  In order to achieve lower range and angle \ac{crb}s, subarrays of centro-symmetric modular architecture should be distributed close to the array edges.
  
\section{Conclusion} \label{conclusion}

In this letter, we proposed a generic modular array architecture and derived the closed-form expressions of range and angle \ac{crb}s based on the \ac{hspm} with distinct \ac{aoa}s.
Numerical results were presented for validating the derived \ac{crb}s and provided the useful insights.
Firstly, the \ac{hspm} with distinct \ac{aoa}s demonstrates the capability to achieve lower range \ac{crb} compared to the \ac{hspm} with shared \ac{aoa}, striking a good balance between modeling accuracy and complexity.
This makes the \ac{hspm} with distinct \ac{aoa}s more suitable for application in the generic modular array architecture.
Additionally, the effectiveness of a non-uniform subarray arrangement was validated, as it could achieve lower range and angle CRB values compared to a traditional uniform subarray arrangement. 
This highlights the potential of the generic modular arrays in near-field sensing. 
For future research, subarray configurations of the generic modular array architecture should be designed  considering practical installation scenarios and joint optimization of communication and sensing capabilities.

\appendices
\section{Proof of Theorem \ref{crb_r_theta}} \label{crb_proof}
	According to \cite{boyer2011performance}, we can obtain the \ac{crb}s for range and angle estimations, which can be given as
	\begin{equation}\label{crb_app}
		\begin{aligned}
			\text{CRB}_r\!=\!\frac{\frac{\sigma _{s}^{2}}{2}\!\left(\!\! {{\left\| {{{\mathbf{\dot{g}}}}_{\theta }} \right\|}^{2}}\!\!-\!\frac{{{\left| \mathbf{\dot{g}}_{\theta }^{H}\mathbf{g} \right|}^{2}}}{{{\left\| \mathbf{g} \right\|}^{2}}}\! \right)}{{{\left| \alpha  \right|}^{2}}\det \left( \mathbf{Q}' \right)},\  
			\text{CRB}_{\theta} \!=\!\frac{\frac{\sigma _{s}^{2}}{2}\!\left(\! \!{{\left\| {{{\mathbf{\dot{g}}}}_{r}} \right\|}^{2}}\!\!-\!\frac{{{\left| \mathbf{\dot{g}}_{r}^{H}\mathbf{g} \right|}^{2}}}{{{\left\| \mathbf{g} \right\|}^{2}}}\! \right)}{{{\left| \alpha  \right|}^{2}}\det \left( \mathbf{Q}' \right)},
		\end{aligned}
	\end{equation}
	where $\mathbf{g}_\theta=\frac{\partial \mathbf{g}}{\partial \theta}$ and $\mathbf{g}_r=\frac{\partial \mathbf{g}}{\partial r}$, and
	\begin{equation}
		\begin{aligned}
			& \det \left( {\mathbf{{Q}'}} \right)=\frac{1}{{{K}^{2}}{{M}^{2}}}\left[ \left( KM{{\left\| {{{\mathbf{\dot{g}}}}_{r}} \right\|}^{2}}-{{\left| \mathbf{\dot{g}}_{r}^{H}\mathbf{g} \right|}^{2}} \right) \right.\times \\ 
			& \left. \left(\! KM{{\left\| {{{\mathbf{\dot{g}}}}_{\theta }} \right\|}^{2}}\!\!-\!{{\left| \mathbf{\dot{g}}_{\theta }^{H}\mathbf{g} \right|}^{2}} \!\right)\!-\!\Re {{\left(\! KM\mathbf{\dot{g}}_{r}^{H}{{{\mathbf{\dot{g}}}}_{\theta }}\!-\!\mathbf{\dot{g}}_{\theta }^{H}\mathbf{g}{{\mathbf{g}}^{H}}{{{\mathbf{\dot{g}}}}_{r}} \!\right)}^{2}}\! \right]. 
		\end{aligned}
	\end{equation}

	According to the the rule for the derivative of the matrix multiplication with respect to the scalar, i.e.,
	\begin{equation}
		\frac{\partial \left( \mathbf{AB} \right)}{\partial x}=\frac{\partial \mathbf{A}}{\partial x}\mathbf{B}+\mathbf{A}\frac{\partial \mathbf{B}}{\partial x},
	\end{equation} 
	we have ${{\mathbf{\dot{g}}}_{u}}=\left( \text{diag}({{\dot{\boldsymbol{\nu}}}_{u}})\otimes {{\mathbf{I}}_{M}} \right)\mathbf{\tilde{a}}+\left( \text{diag}\left(  \boldsymbol{\nu} \right)\otimes {{\mathbf{I}}_{M}} \right)({{\mathbf{\dot{\tilde{a}}}}_{u}}), u\in\left\{r,\theta\right\}$.
	By denoting ${{\dot{\boldsymbol{\nu}}}_{u}}\triangleq\frac{\partial \boldsymbol{\nu} }{\partial u}$ and ${{\mathbf{\dot{\tilde{a}}}}_{u}}\triangleq \frac{\partial \mathbf{\tilde{a}}}{\partial u}$ for all $u\in\left\{r,\theta\right\}$, we can derive the following results:
		\begin{equation} \label{relevant_terms1}
		\begin{aligned}
			&{{{\mathbf{\dot{g}}}}_{v}}^{H}{{{\mathbf{\dot{g}}}}_{u}}\\ 
			&\overset{\left(a\right)}{=}\!{{{\mathbf{\tilde{a}}}}^{H}}\!\!\left(\! \text{diag}\!\left(\! {{\dot{{\boldsymbol{\nu}}}}}_{v}^{H} \!\right)\!\text{diag}\!\left( \!{{{\dot{{\boldsymbol{\nu}} }}}_{u}}\! \right)\!\otimes\! {{\mathbf{I}}_{M}}\! \right)\mathbf{\tilde{a}}\!+\!{{{\mathbf{\tilde{a}}}}^{H}}\!\left(\! \text{diag}\!\left(\! {\dot{{\boldsymbol{\nu}}}}_{v}^{H}\! \right)\!\text{diag}\!\left( \!{\boldsymbol{\nu}} \! \right)\!\otimes \!{{\mathbf{I}}_{M}} \!\right)\!{{{\mathbf{\dot{\tilde{a}}}}}_{u}} \\ 
			& +\mathbf{\dot{\tilde{a}}}_{v}^{H}\!\!\left(\! \text{diag}\!\left(\! {{{\boldsymbol{\nu}} }^{H}}\! \right)\!\text{diag}\!\left( \!{{{\dot{{\boldsymbol{\nu}} }}}_{u}}\! \right)\!\otimes\! {{\mathbf{I}}_{M}} \!\right)\!\mathbf{\tilde{a}}\!+\!\mathbf{\dot{\tilde{a}}}_{v}^{H}\!\!\left( \!\text{diag}\!\left( {{{\boldsymbol{\nu}} }^{H}} \!\right)\!\text{diag}\!\left( {\boldsymbol{\nu}}  \right)\!\otimes\! {{\mathbf{I}}_{M}} \!\right)\!{{{\mathbf{\dot{\tilde{a}}}}}_{u}},	
		\end{aligned}
	\end{equation}	
	where $vu \in\left\{rr,r\theta,\theta\theta\right\}$, and the derivation of $(a)$ is based on the property of the Kronecker product $\left( \mathbf{A}\otimes \mathbf{B} \right)\left( \mathbf{C}\otimes \mathbf{D} \right)=\left( \mathbf{AC}\otimes \mathbf{BD} \right)$.
	Similarly, we can obtain that
	\begin{equation}\label{relevant_terms2}
		\begin{aligned}
			{{{\mathbf{\dot{g}}}}_{u}}^{H}\mathbf{g} \!=\!{\mathbf{\tilde{a}}^{H}}\!\!\left(\!\text{diag}\!\left(\! {{{\dot{\boldsymbol{\nu} }}}_{u}}^{H}\! \right)\!\text{diag}\!\left(\! \boldsymbol{\nu} \!\right)\!\otimes\! {{\mathbf{I}}_{M}} \!\right)\!\mathbf{\tilde{a}}
			\!+\!{{{\mathbf{\dot{\tilde{a}}}}}_{u}^{H}}\!\!\left(\! \text{diag}\!\left(\! {{\boldsymbol{\nu}}^{H}}\! \right)\!\text{diag}\!\left(\! \boldsymbol{\nu}\!  \right)\!\otimes\! {{\mathbf{I}}_{M}}\! \right)\!\mathbf{\tilde{a}}, 
		\end{aligned}
	\end{equation}
	where $u\in \left\{r,\theta\right\}$.
	Furthermore, we can obtain that 
	\begin{subequations} \label{element_derivation}
		\begin{align}
			&{{\mathbf{\dot{\tilde{a}}}}_{u}}=\frac{\partial {{[{{({{\mathbf{a}}_{1}})}^{T}},\ldots ,{{({{\mathbf{a}}_{K}})}^{T}}]}^{T}}}{\partial u}={{\left[ \frac{\partial \mathbf{a}_{1}^{T}}{\partial u},\ldots ,\frac{\partial \mathbf{a}_{K}^{T}}{\partial u} \right]}^{T}},\\
			&{{\left[ \frac{\partial \mathbf{a}_{k}^{T}}{\partial u} \right]}_{m}}\!\!\!
			=\!-j\frac{2\pi md}{\lambda }\frac{\partial \!\sin\!{{\theta}_{k}}}{\partial u}{{e}^{-j\!\frac{2\pi }{\lambda}\!md\!\sin\!{{\theta }_{k}}}},\\
			&{{\left[ {{{\dot{\boldsymbol{\nu} }}}_{u}} \right]}_{k}}=-j\frac{2\pi }{\lambda }\frac{\partial {{r}_{k}}}{\partial u}{{e}^{-j\frac{2\pi }{\lambda }{{r}_{k}}}},
		\end{align}
	\end{subequations}
	where $	{{\left[ \frac{\partial \mathbf{a}_{k}^{T}}{\partial u} \right]}_{m}}$ is the $m$-th element of $ \frac{\partial \mathbf{a}_{k}^{T}}{\partial u} $, and ${{\left[ {{{\dot{\boldsymbol{\nu} }}}_{u}} \right]}_{k}}$ is the $k$-th element of ${{{\dot{\boldsymbol{\nu} }}}_{u}}$.
	
	Based on (\ref{element_derivation}) and (\ref{theorem1_inter}), we can derive the expressions constituting the terms of $	{{{\mathbf{\dot{g}}}}_{v}}^{H}{{{\mathbf{\dot{g}}}}_{u}}$, $ vu \in\left\{rr,\theta\theta,r\theta\right\}$, which can be given by
	\begin{equation} \label{constitute_termsaa1}
		\begin{aligned}
				&	{{{\mathbf{\tilde{a}}}}^{H}}\left( \text{diag}\left( {{{\dot{\boldsymbol{\nu} }}}_{v}}^{H} \right)\text{diag}\left( {{{\dot{\boldsymbol{\nu} }}}_{u}} \right)\otimes {{\mathbf{I}}_{M}} \right)\mathbf{\tilde{a}}=M\sum\limits_{k\in\mathcal{K}}{\left[ {{{\dot{\boldsymbol{\nu} }}}_{v}} \right]_{k}^{*}{{\left[ {{{\dot{\boldsymbol{\nu} }}}_{u}} \right]}_{k}}}\\ 
			&=\!M\!{{\left(\!\frac{2\pi }{\lambda }\!\right)}^{2}}\!\!\!\sum\limits_{k\in\mathcal{K}}{\!\!\left(\! \frac{\partial {{r}_{k}}}{\partial v} \!\right)\!\!\left(\!\frac{\partial {{r}_{k}}}{\partial u} \!\right)}\!=\!M\!{{\left(\! \frac{2\pi }{\lambda }\!\right)}^{2}}\!\!\Upsilon  
			,\Upsilon \!\!=\!\!\left\{\!\! \begin{array}{*{35}{l}}
				q, & vu=rr  \\
				\tilde{q}, &  vu=\theta\theta  \\
				\hat{q}, &  vu=r\theta  \\
			\end{array}, \right.
		\end{aligned}
	\end{equation}
	\begin{equation} \label{constitute_termsaa2}
	\begin{aligned}
		&{{ {{{\mathbf{\dot{\tilde{a}}}}}_{v}^{H}}}}\left( \text{diag}\left( {{\boldsymbol{\nu} }^{H}} \right)\text{diag}\left( {{{\dot{\boldsymbol{\nu} }}}_{u}} \right)\otimes {{\mathbf{I}}_{M}} \right)\mathbf{\tilde{a}}\\
        &=\!\sum\limits_{k\in\mathcal{K}}{\!\left[ \boldsymbol{\nu} \right]_{k}^{*}{{\left[ {{{\dot{\boldsymbol{\nu} }}}_{u}} \right]}_{k}}}\!\sum\limits_{m\in\mathcal{M}}\!{\!\left[ \frac{\partial {{\mathbf{a}}_{k}}}{\partial v} \right]_{m}^{\text{*}}}\!{{\!\left[ {{\mathbf{a}}_{k}} \right]}_{m}} \\ 
		& =\sum\limits_{k\in\mathcal{K}}{\left( -j\frac{2\pi }{\lambda } \right)\left( \frac{\partial {{r}_{k}}}{\partial u} \right)}\sum\nolimits_{m=-\frac{M-1}{2}}^{\frac{M-1}{2}}{\left( j\frac{2\pi }{\lambda }md\frac{\partial \sin {{\theta }_{k}}}{\partial v} \right)} \\ 
		& =\!d{{\left( \!\frac{2\pi }{\lambda } \!\right)}^{2}}\!\!\sum\limits_{k\in\mathcal{K}}\!\!{\left(\! \frac{\partial {{r}_{k}}}{\partial u} \!\right)}\!\!\left( \frac{\partial \sin {{\theta }_{k}}}{\partial v} \right)\sum\nolimits_{m=-\frac{M-1}{2}}^{\frac{M-1}{2}}{m}= 0,
	\end{aligned}
\end{equation}
\begin{equation}\label{constitute_termsaa3}
	\begin{aligned}
	&\mathbf{\dot{\tilde{a}}}_{v}^{H}\left( \text{diag}\left( {{\boldsymbol{\nu} }^{H}}\right)\text{diag}\left( \boldsymbol{\nu}  \right)\otimes {{\mathbf{I}}_{M}} \right){{{\mathbf{\dot{\tilde{a}}}}}_{u}}\\
		& ={{\left( \frac{2\pi d}{\lambda } \right)}^{2}}\sum\limits_{k\in\mathcal{K}}{\left( \frac{\partial \sin {{\theta }_{k}}}{\partial v} \right)\left( \frac{\partial \sin {{\theta }_{k}}}{\partial u} \right)\sum\limits_{m=-\frac{M-1}{2}}^{\frac{M-1}{2}}{{{m}^{2}}}} \\ 
		& =\frac{{{M}^{3}}-M}{12}{{\left( \frac{2\pi d}{\lambda } \right)}^{2}}\sum\limits_{k\in\mathcal{K}}{\left( \frac{\partial \sin {{\theta }_{k}}}{\partial v} \right)\left( \frac{\partial \sin {{\theta }_{k}}}{\partial u} \right)}\\  
		& \overset{\left(a\right)}{=}\frac{{{M}^{3}}-M}{12}{{\left( \frac{2\pi d}{\lambda } \right)}^{2}}\Xi, \Xi=\left\{ \begin{array}{*{35}{l}}
			z, & vu=rr  \\
			\tilde{z}, &  vu=\theta\theta  \\
			\hat{z}, &  vu=r\theta  \\
		\end{array}, \right. 	
	\end{aligned}
\end{equation}
	where the derivation of $(a)$ is based on the fact that
$\sum\nolimits_{m=-\frac{M-1}{2}}^{\frac{M-1}{2}}{{{m}^{2}}}=\frac{{{M}^{3}}-M}{12}$.
	Similarly, we can deduce the expressions constituting the terms of ${{{\mathbf{\dot{g}}}}_{u}}^{H}\mathbf{g}$, $u\in\left\{r,\theta\right\}$ as
		\begin{equation} \label{constitute_terms22}
		\begin{aligned}
			&{{{\mathbf{\tilde{a}}}}^{H}}\left( \text{diag}\left( {{{\dot{\boldsymbol{\nu} }}}_{u}}^{H} \right)\text{diag}\left( \boldsymbol{\nu}  \right)\otimes {{\mathbf{I}}_{M}} \right)\mathbf{\tilde{a}}=j\frac{2\pi M}{\lambda }\Omega , \Omega  =\! \left\{ \!\begin{array}{*{35}{l}}
				p, & u=r  \\
				\tilde{p}, & u=\theta  \\
			\end{array}, \right.\\
			&	{{{{{\mathbf{\dot{\tilde{a}}}}}_{u}^{H}} }}\left( \text{diag}\left({{\boldsymbol{\nu} }^{H}} \right)\text{diag}\left( \boldsymbol{\nu}  \right)\otimes {{\mathbf{I}}_{M}} \right)\mathbf{\tilde{a}}=0.		  
		\end{aligned}
	\end{equation}

	By substituting (\ref{constitute_termsaa1}),  (\ref{constitute_termsaa2}),  (\ref{constitute_termsaa3}) and (\ref{constitute_terms22}) into (\ref{relevant_terms1}) and (\ref{relevant_terms2}), we can obtain the expressions for the relevant terms as follows:
	\begin{equation}\label{crb_term_gg11}
		\begin{aligned}
			{{{\mathbf{\dot{g}}}}_{v}}^{H}{{{\mathbf{\dot{g}}}}_{u}}&={{\left( \frac{2\pi }{\lambda } \right)}^{2}}\frac{\left( {{M}^{3}}-M \right){{d}^{2}}\Xi +12M\Upsilon  }{12},\\
			&\left\{ \begin{array}{*{35}{l}}
				\Upsilon =q,\Xi=z, & \textit{if}\ vu=rr  \\
				\Upsilon =\tilde{q},\Xi=\tilde{z}, & \textit{if}\   vu=\theta\theta  \\
				\Upsilon =\hat{q},\Xi=\hat{z}, & \textit{if}\   vu=r\theta  \\
			\end{array}, \right. \\
			{{{\mathbf{\dot{g}}}}_{u}}^{H}\mathbf{g}&=j\frac{2\pi M}{\lambda }\Omega ,\Omega =\left\{ \begin{array}{*{35}{l}}
				p, & u=r  \\
				\tilde{p}, & u=\theta   \\
			\end{array}. \right.  
		\end{aligned}
	\end{equation}
	By substituting (\ref{crb_term_gg11}) into (\ref{crb_app}), we can obtain the closed-form of range/angle \ac{crb} in (\ref{crb_closedform}), which completes the proof.

\section{Proof of Corollary \ref{crb_specialcase}} \label{corollary_1_proof}
When $\theta=0$, we have $\sin\theta_k=\frac{-x_k}{r_k}$, $\cos\theta_k=\frac{r}{r_k}$, and $r_k=\sqrt{r^2+x_k^2}$. Substituting these expressions into the intermediate variables in Theorem \ref{crb_r_theta}, we can obtain the expression of $p'$, $q'$, $z'$, $\tilde{q}'$, $\tilde{z}'$.
According to the geometric properties of centro-symmetric arrays, it can be inferred that 
\begin{equation}
\begin{aligned}
\tilde{p}\!=\!\!\sum_{k\in\mathcal{K}}\frac{-rx_k}{r^2+s_k^2}\! =\! 0,
    \hat{q}\!=\!\!\sum_{k\in\mathcal{K}}\frac{-r^2x_k}{r^2+x_k^2}\!=\! 0,
    \hat{z}\!=\!\! \sum_{k\in\mathcal{K}}\frac{r^4x_k}{r^2+x_k^2}\!=\!0.  
\end{aligned} 
\end{equation}

By substituting these simplified expressions into \eqref{crb_r_hspm_dis}  and \eqref{crb_theta_hspm_dis}, we arrive at the simplified CRB expressions in \eqref{crb_r_hspm_dis_special} and \eqref{crb_theta_hspm_dis_special}.

	\bibliographystyle{IEEEtran}
	\bibliography{mybib}
\end{document}